\documentclass[prb,notitlepage,showpacs,onecolumn,aps,superscriptaddress,a4paper]{revtex4-1}

\usepackage{color}
\usepackage{dcolumn}
\usepackage{amsmath}
\usepackage{amssymb}
\usepackage{mathtools}
\usepackage{graphicx}
\usepackage{latexsym}
\usepackage{amsfonts}
\usepackage{amssymb}
\usepackage{comment}
\usepackage{natbib}
\DeclareGraphicsExtensions{.pdf,.gif,.jpg}

\def\hc{\text{h.c.}}
\def\a{\alpha}
\def\eps{\epsilon}

\begin{document}


\title{Electron traversal times in disordered graphene nanoribbons}

\author{Michael Ridley}
\affiliation{The Raymond and Beverley Sackler Center for Computational Molecular and Materials Science, Tel Aviv University, Tel Aviv 6997801, Israel}

\author{Michael A. Sentef}
\affiliation{Max Planck Institute for the Structure and Dynamics of Matter, 22761 Hamburg, Germany}

\author{Riku Tuovinen}
\email{riku.tuovinen@mpsd.mpg.de}
\affiliation{Max Planck Institute for the Structure and Dynamics of Matter, 22761 Hamburg, Germany}


\begin{abstract}
Using the partition-free time-dependent Landauer-B{\"u}ttiker formalism for transient current correlations, we study the traversal times taken for electrons to cross graphene nanoribbon (GNR) molecular junctions. We demonstrate electron traversal signatures that vary with disorder and orientation of the GNR. These findings can be related to operational frequencies of GNR-based devices and their consequent rational design.
\end{abstract}

\maketitle


\section{Introduction}

A fundamental property limiting the operational frequency of a molecular device is the traversal time $\tau_{\text{tr}}$ for electronic information to cross between the nanojunction terminals~\cite{Dragoman2011}. For instance, in graphene, the cutoff frequency $f_{\text{max}}$ is related to the traversal time as $f_{\text{max}}=1/2\pi\tau_{\text{tr}}$~\cite{Lin2009,Liao2010}. For the molecular electrician, this raises the key question of how long it takes for electronic information to propagate across a nanosized device, as this sets a fundamental limit on the speed of the device operation. In quantum mechanics, time does not have the same status as a dynamical variable such as the energy or particle position. In fact much debate has centered around the correct definition of the traversal time through a generic potential barrier~\cite{Buttiker1982,Hauge1989}, as well as the relation of this quantity to the dwell time (time spent in the molecular region)~\cite{Collins1987}, the Larmor clock time (the time taken to move between scattering channels)~\cite{Baz1967b,Rybachenko} and the group delay time (the time delay in the nonlocal propagating wave packet caused by scattering off the potential barrier)~\cite{Winful2006}. 
Crucially, all the aforementioned times are defined in terms of the transmission probability, potential and incident energy of electrons moving in a static scattering theory picture~\cite{Landsman2015}, so that a theory which takes strong time-dependence into account is still needed. This is crucial for the understanding of laser-stimulated tunneling processes and related to the problem of tunneling times in strong field ionization experiments~\cite{Landsman2014,Camus2017,Hofmann2019}.

Graphene nanoribbon (GNR)-based molecular junctions are excellent candidates for room-temperature transistors, i.e., graphene field-effect transistors (GFETs)~\cite{Lin2009}, GHz-THz frequency modulators~\cite{Gao2014}, and photodetectors~\cite{Koppens2014}, due to their high mobility and charge carrier saturation velocities. GNRs can be engineered with band gaps that are tunable via the nanoribbon symmetry properties and widths~\cite{Chen2015}, and currently, sub-$10$~nm nanoribbon widths are accessible from chemical fabrication techniques~\cite{Li2008,Kimouche2015,Carbonell2018,Li2019a,Li2019b}. Recent experimental progress shows an inverse scaling of operational frequency with the nanoribbon length~\cite{Liao2010}, or with the square of the nanoribbon length~\cite{Li2008} for ribbons whose carrier drift velocity scales inversely with ribbon length. Typically the maximum operational frequencies of radio frequency (RF) GFETs exceed those of Silicon-based transistors with the same dimensions~\cite{Liao2010}, and can in principle be achieved in the $100$~GHz range~\cite{Lin2010,Wu2011,Cheng2012}. The cutoff frequencies of GFETs are strongly affected by the presence of defects and flexibility in the nanoribbon, but $100$~GHz flexible nanoribbons have also recently been fabricated~\cite{Yu2018}.

Disorder may have a profound impact on the operation of the graphene-based devices. For example, it has been investigated that edge disorder affects the armchair-oriented GNR (AGNR) more than the zigzag-oriented ones (ZGNR)~\cite{Areshkin2007,Dauber2014}. This is because the edge states in ZGNRs are energetically protected against impurity perturbations. This is not the case for AGNRs, in which the edge states are less dominant so that disorder has a much larger effect on the conductance~\cite{Mucciolo2009,Mucciolo2010}. On the other hand, disorder-induced broken chiral symmetry in terms of random bond disorder was considered in Ref.~\cite{Zhu2016}.

In this paper we expand upon a recent proposal~\cite{Ridley2017} to investigate traversal times for electrons moving in disordered GNRs by looking directly at the dynamics of statistical correlations between electronic signals measured in different reservoirs connected to the GNR. 
We demonstrate that the traversal time has a clearer signature in AGNR than in ZGNR. This is because the charge densities in AGNR structures are more delocalised than in ZGNR, where the formation of standing-edge-state charge waves leads to wave fronts with a spatial orientation lying diagonal across the plane of the nanoribbon~\cite{Tuovinen2014,daRocha2015,Tuovinen2019b}. We also show that the effect of breaking chiral symmetry (on-site disorder) has less effect on the traversal times compared with the disorder that preserves chirality (hopping disorder)\footnote{A random on-site potential breaks the sublattice symmetry and can broaden possible Landau levels, i.e. the chiral symmetry is destroyed. This does not happen with nearest neighbor hopping disorder as hopping between different sublattices is by construction the same in both directions~\cite{Ludwig1994,Kawarabayashi2009,Zhu2016,Chen2018}.}.

\section{Model and method}

\begin{figure}[t]
\centering
\includegraphics[width=0.65\textwidth]{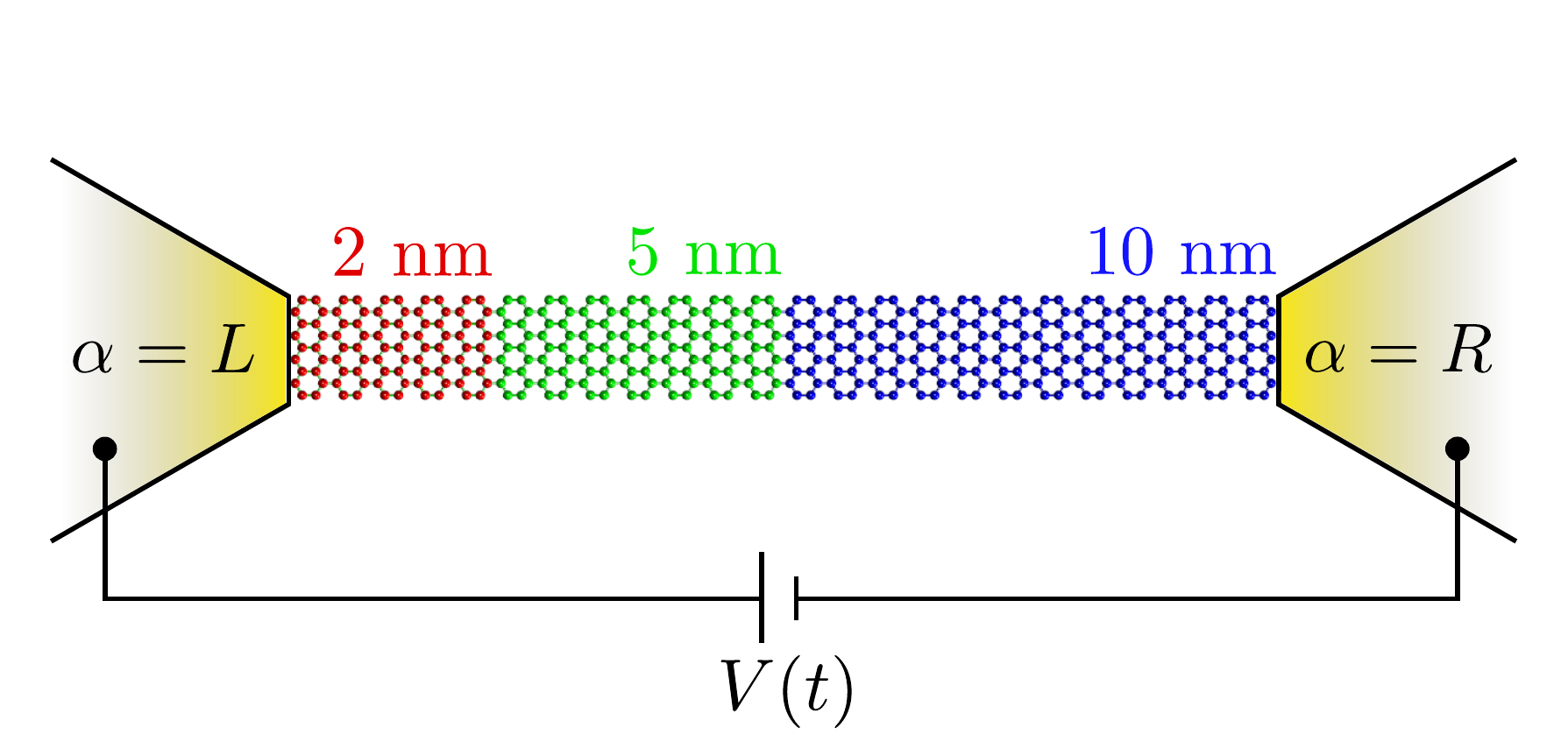}
\caption{Transport setup for an armchair graphene nanoribbon where the left-most carbon atoms are connected to the $\alpha=L$ lead and the right-most carbon atoms are connected to the $\alpha=R$ lead. For times $t\geq 0$, a voltage bias $V(t)$ is applied in the leads and charge carriers start to flow through the graphene junction. We consider ribbons of varying lengths ($2$, $5$, and $10$~nm), and similarly also the zigzag orientation (not shown).}
\label{fig:setup}
\end{figure}

Our transport setup (cf. Figure~\ref{fig:setup}) is described by the Hamiltonian $\hat{H} = \hat{H}_{\text{lead}} + \hat{H}_{\text{lead-GNR}} + \hat{H}_{\text{GNR}}$ where the lead environment is given by
\begin{equation}
\hat{H}_{\text{lead}} = \sum_{k\alpha} \eps_{k\alpha} \hat{c}_{k\alpha}^\dagger \hat{c}_{k\alpha}.
\end{equation}
At the initial switch-on time $t_{0}$ the energy dispersion is shifted by the bias voltage $\eps_{k\alpha} \to \eps_{k\alpha} + V_\alpha\left(t\right)$. The left-most atoms are connected to the left lead ($L$) whereas the right-most atoms are connected to the right lead ($R$). The coupling between the leads ($\a = L,R$) and the GNR has the form
\begin{equation}
\hat{H}_{\text{lead-GNR}} = \sum_{mk\alpha} (T_{mk\alpha} \hat{c}_m^\dagger \hat{c}_{k\alpha} + \hc),
\end{equation}
where we wish to emphasize that the coupling matrix elements $T_{mk\a}$ are kept constant for all times (i.e., we work within the partition-free approach~\cite{Cini1980,Stefanucci2004,Ridley2018}). In practice, we choose the lead bandwidth, given by the energy dispersions $\eps_{k\a}$, to be much larger than the coupling energies so that we may employ the wide-band limit approximation (WBLA), in which the level width matrix ${\varGamma}_{\a,mn}\left(\omega\right)=2 \pi \sum_k T_{mk\a} \delta(\omega-\epsilon_{k\a}) T_{k\a n}$ is evaluated at the Fermi energy of lead $\a$. The WBLA is justified because we are interested in a regime where the lead-GNR coupling is weaker than the internal hopping within the GNR~\cite{Zhu2005,Verzijl2013,Covito2018,Ridley2019}, as this also enables us to focus on the effect on traversal time caused by the internal ribbon structure. 
The GNR is modeled by a single $\pi$-orbital tight-binding picture
\begin{equation}
\hat{H}_{\text{GNR}} = \sum_{mn} h_{mn} \hat{c}_m^\dagger \hat{c}_n ,
\end{equation}
where the intramolecular hopping parameters $h_{mn}$ are nonzero for the nearest neighbors only, and set by the typically used carbon-carbon hopping integral in graphene $h_{mn}=t_C=2{.}7$~eV~\cite{Reich2002,CastroNeto2009,Harju,Joost2019}. Longer-range hoppings could be included in the model similarly but here we wish to preserve the particle-hole symmetry of the undisordered GNR.

We employ the recently developed time-dependent Landauer-B{\"u}ttiker (TD-LB) formalism~\cite{Landauer1970,Buttiker1986,Tuovinen2013,Tuovinen2014,Ridley2015,Ridley2016a,Ridley2016b,Tuovinen2016a,Tuovinen2016b,Tuovinen2019a} to compute the two-time current correlation function $C_{\alpha\beta}\left(t_{1},t_{2}\right)\equiv\left\langle \Delta\hat{I}_{\alpha}\left(t_{1}\right)\Delta\hat{I}_{\beta}\left(t_{2}\right)\right\rangle$ between the current deviation operators $\Delta\hat{I}_{\alpha}\left(t_{1}\right)\equiv\hat{I}_{\alpha}\left(t_{1}\right)-\left\langle \hat{I}_{\alpha}\left(t_{1}\right)\right\rangle$ between different leads labelled by $\alpha$ and $\beta$. The current operator of lead $\alpha$ is related to the particle number there via $\hat{I}_{\alpha}\left(t\right)\equiv qd\hat{N}_{\alpha}/dt$, where $q$ is the charge of the particle. When there is no variation of current in one of the leads, i.e. $\Delta\hat{I}_{\alpha}\left(t_{1}\right)=0$ the correlation between this signal and the current variation in the other leads is trivially zero.

In a two-terminal junction (cf. Figure~\ref{fig:setup}), the labels $\alpha$ and $\beta$ can refer to either the left ($L$) or right ($R$) terminals, whose energies are shifted symmetrically to create a voltage drop of $V\left(t\right)$ across the system $V\left(t\right)/2=V_{L}\left(t\right)=-V_{R}\left(t\right)$. We note here that the driving bias voltage in our setup can be of dc type (sudden quench) or ac type (time-dependent modulation)~\cite{Tuovinen2014,Ridley2015,Ridley2017b}. In Ref.~\cite{Ridley2017} it was shown that timescales associated with electron traversal times and internal reflection processes could be seen as resonances in the real part of symmetrized cross-lead correlations $C^{\left(\times\right)}\left(t+\tau,t\right)\equiv\left[C_{LR}\left(t+\tau,t\right)+C_{RL}\left(t+\tau,t\right)\right]/2$, as a function of the relative time $\tau$. The traversal time $\tau_{\text{tr}}$ is therefore defined by the following relation:
\begin{equation}
\textrm{max}\left|\textrm{Re}\left[C^{\left(\times\right)}\left(t+\tau,t\right)\right]\right|\equiv\left|\textrm{Re}\left[C^{\left(\times\right)}\left(t\pm\tau_{\text{tr}},t\right)\right]\right|\label{eq:ttime_def} .
\end{equation}

Within the WBLA, the cross-correlations are evaluated in terms of Keldysh components of the Green's function ${G}\left(t_{1},t_{2}\right)$ (projected onto the GNR subspace)~\cite{Ridley2017}
\begin{eqnarray}
C^{\left(\times\right)}\left(t_{1},t_{2}\right) & = & 2q^{2}\underset{\a\neq\a'}{\sum}\mathrm{Tr}_{\text{GNR}}\Big\{{\varGamma}_{\a}{G}^{>}\left(t_{1},t_{2}\right){\varGamma}_{\a'}{G}^{<}\left(t_{2},t_{1}\right)\nonumber \\
& + & \mathrm{i}{G}^{>}\left(t_{1},t_{2}\right)\big[{\varLambda}_{\a}^{+}\left(t_{2},t_{1}\right){\varGamma}_{\a'}+{\varGamma}_{\a}\large({\varLambda}_{\a'}^{+}\large)^{\dagger}\left(t_{1},t_{2}\right)\big]\nonumber\\
& + & \mathrm{i}\big[{\varLambda}_{\a}^{-}\left(t_{1},t_{2}\right){\varGamma}_{\a'}+{\varGamma}_{\a}\large({\varLambda}_{\a'}^{-}\large)^{\dagger}\left(t_{2},t_{1}\right)\big]{G}^{<}\left(t_{2},t_{1}\right)\nonumber\\
& - & {\varLambda}_{\a}^{+}\left(t_{2},t_{1}\right){\varLambda}_{\a'}^{-}\left(t_{1},t_{2}\right)-\large({\varLambda}_{\a'}^{+}\large)^{\dagger}\left(t_{1},t_{2}\right)\large({\varLambda}_{\a}^{-}\large)^{\dagger}\left(t_{2},t_{1}\right)\Big\}\label{eq:correlator},
\end{eqnarray}
where the ${\varLambda}_{\a}^{\pm}\left(t_{1},t_{2}\right)$ matrices are defined in terms of convolutions on the real and imaginary branches on the complex time contour~\cite{Ridley2017}. We will investigate the dynamics of steady state (the switch-on time is taken to $t_{0}\rightarrow-\infty$) cross-correlations, which are accessible experimentally~\cite{fevrier2018}.

We note that there is some spreading in the individual resonant peaks associated with traversal times in the correlator, so that our proposal takes into account the probabilistic nature of electron propagation in accordance with realistic proposals for traversal time distributions~\cite{Fertig1990,Landsman2014,Landsman2015}. This is particularly relevant for our approach which enables us to study arbitrary time-dependent biases, e.g., in which the drive is stochastic in time~\cite{Ridley2016b}. The Fourier transform of the real part of $C^{\left(\times\right)}\left(t+\tau,t\right)$, with respect to the relative time $\tau$, is equivalent to the frequency-dependent power spectrum associated with cross-lead correlations: 
\begin{equation}
\underset{t_{0}\rightarrow-\infty}{\lim}\mathcal{F}\left\{\textrm{Re}\left[C^{\left(\times\right)}\left(t+\tau,t\right)\right];\omega\right\}=\frac{P_{LR}\left(\omega\right)+P_{RL}\left(\omega\right)}{2}\label{eq:spectrum_cross} .
\end{equation}
Here, $P_{\alpha\beta}\left(\omega\right)$ is defined as the Fourier transform with respect to $\tau$ of the real part of $C_{\alpha\beta}\left(t+\tau,t\right)$~\cite{Ridley2017}. In practice, the high-frequency component of the current fluctuations can be probed by studying the infrared-to-optical frequency range of light emitted by the junction~\cite{fevrier2018}.
 
The central idea of our work is that one should quantify the traversal time for electronic information to cross the system by looking directly at the correlations in the electronic signal itself, rather than trying to build an indirect definition of operational time from the calculation of transmission probabilities. The definition of traversal time here is closely related to the definition of Miller and Pollak, which makes use of flux-flux correlation functions~\cite{Pollak1984}. However, the TD-LB formalism is valid for arbitrary lead temperatures, lead-GNR hybridization strengths, and time-dependent biases. 

\section{Results and discussion}

As we consider the WBLA, the detailed electronic structure of the leads is not important for the description of the transport properties of the GNR. We then fix the coupling strength between the GNR and the leads by the frequency-independent resonance width $\varGamma_\a = t_C/10$ corresponding to a weak-coupling regime where the WBLA is a good approximation~\cite{Zhu2005,Verzijl2013,Covito2018,Ridley2019}. This is further justified in typical transport setups where the bandwidth of the leads is sufficiently large (e.g. gold electrodes) compared to the applied bias voltage. As we wish to preserve the charge neutrality of the GNR in equilibrium, we set the chemical potential to $\mu=0$. The global equilibrium temperature is set by $(k_{\text{B}} T)^{-1} = 100 t_C^{-1}$ ($T$ $=$ 313 K).

\begin{figure}[t]
\centering
\includegraphics[width=\textwidth]{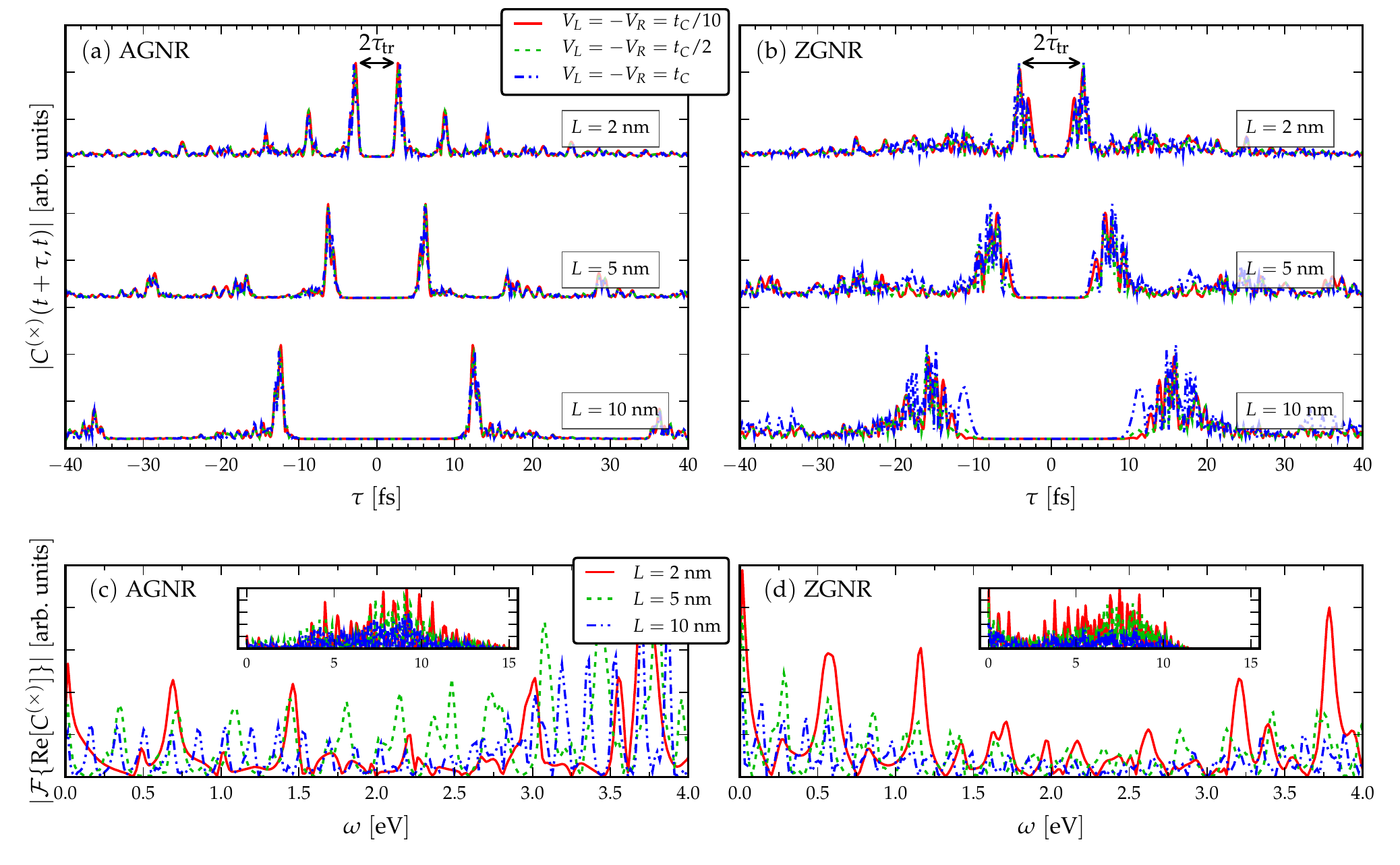}
\caption{Absolute value of the current cross-correlation $C^{\left(\times\right)}\left(t+\tau,t\right)$ at long observation times $t\rightarrow\infty$ for various undisordered GNR samples with varying bias voltage and the consequent Fourier spectra. (a) AGNR of various lengths, (b) ZGNR of various lengths, (c) The low-frequency region of the Fourier transform of panel (a) for fixed bias voltage $V_L=-V_R=t_C/2$ (the inset shows the full frequency range), (d) The low-frequency region of the Fourier transform of panel (b) for fixed bias voltage $V_L=-V_R=t_C/2$ (the inset shows the full frequency range).}
\label{fig:nodisorder}
\end{figure}

\subsection{Response to a dc drive}

It is instructive to first study the current correlations in GNRs without disorder. Figure~\ref{fig:nodisorder} shows the current cross-correlations of undisordered AGNRs and ZGNRs of various lengths and \emph{time-independent} bias voltages. We can make many general observations from the data:
\begin{itemize}
    \item The signal is more clear for the AGNR than ZGNR. In the AGNR case, the propagating wavefront is coherent~\cite{Tuovinen2014}, so that there is less spread in the resonant traversal time signal than in the corresponding ZGNR case. This relates to the shape of the propagating wavefront, since in AGNR it is flat whereas in ZGNR it has a triangular shape~\cite{Tuovinen2014}. The back-and-forth internal reflections of the wavepackets between the electrode interfaces have a fairly regular structure in AGNR which results in a clear signal in the current cross-correlation. This means devices based on ZGNR have a less well-defined operational frequency.
    \item The current cross-correlations are mostly independent of the strength of the applied voltage. The voltage may affect the shape of the curves slightly, but not the location of the main resonance. This can be related to the group velocity of electrons crossing the GNR, $v_k=d\epsilon_k/dk$, which should not depend on a $k$-independent shift in the energy dispersion $\epsilon_k$~\cite{Ridley2017}.
    \item Evidently there is a roughly linear increase of the time-difference between the first maxima with increasing $L$, due to the time taken for the propagating electron wavefront to cross the structure. The time-difference between the first maxima $\tau_{\mathrm{max}}$ is related to the traversal time of information through the GNRs via Eq. \eqref{eq:ttime_def}.
    \item Increasing the length in the AGNR does not increase the number of resonant peaks in the cross-correlations, but in ZGNR it leads to a broader range of resonances clustered about a mean traversal time. This dependence on the orientation of the GNR then affects the spread of operational device frequencies.
    \item The low-frequency regions of the Fourier transforms show resonant frequencies at $\omega=n\varOmega_L$ where $n$ is a positive integer and $\varOmega_L$ is some intrinsic frequency depending on the length of the GNR. In particular, by increasing the length of the GNR more transport channels are opened in the bias window, and therefore more peaks appear in the Fourier spectra.
    \item From the full frequency ranges of the Fourier transforms (insets) we observe a high-frequency operational cut-off which is smaller for the ZGNR case than for the AGNR case. This is itself an interesting effect, as it sets a limit on switches built with these kind of nanoribbons~\cite{Lin2009,Liao2010}.
\end{itemize}

\begin{figure}[t]
\centering
\includegraphics[width=\textwidth]{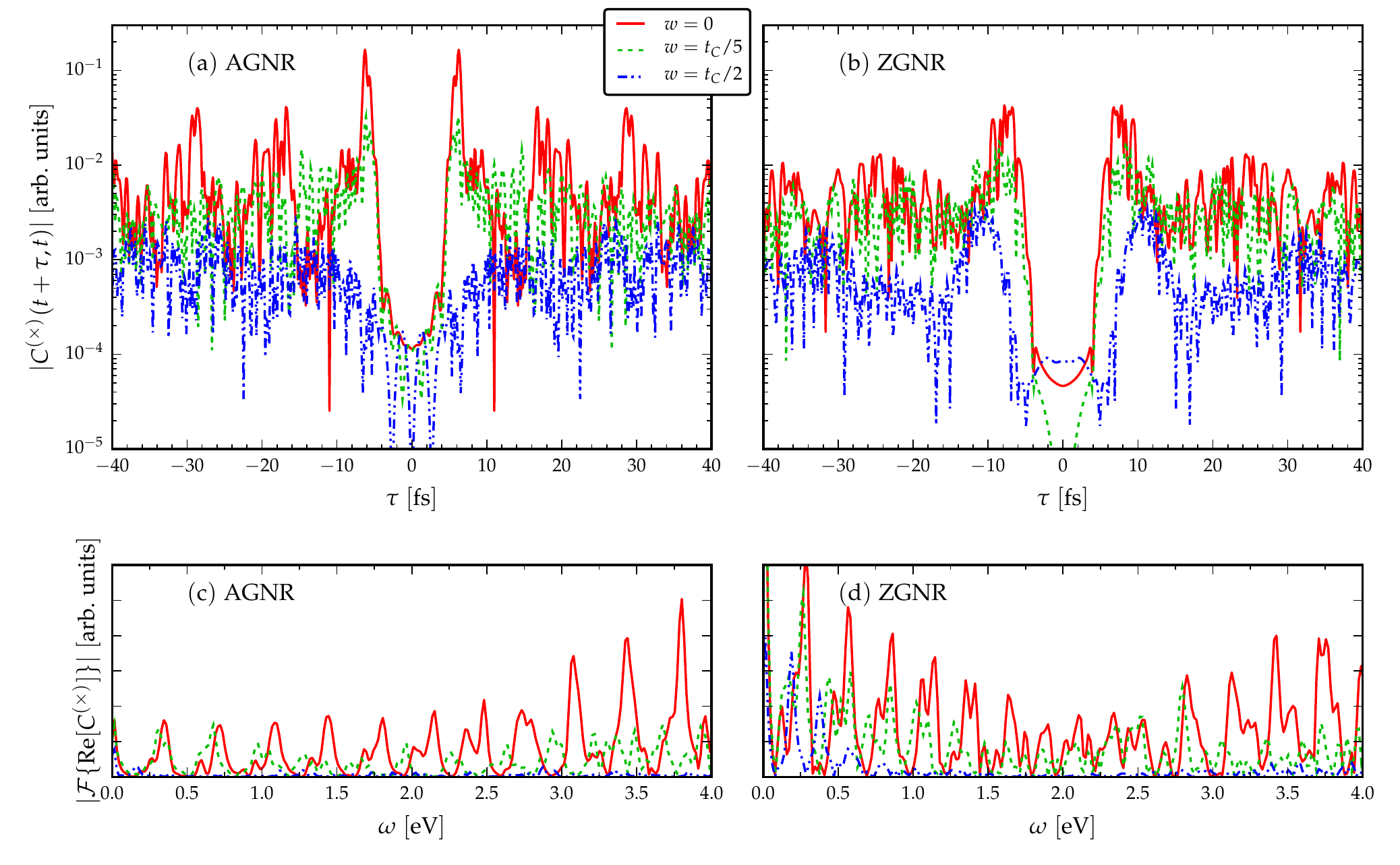}
\caption{Current cross-correlations for fixed length disordered GNRs ($L=5$~nm) with fixed voltage $V_L=-V_R=t_C/2$. A uniformly-distributed disorder $w$ is included in the intramolecular hopping energies, which preserves the chiral symmetry.}
\label{fig:hopdisorder}
\end{figure}

\subsection{The role of disorder}

As we have now established general principles for the traversal times, we concentrate our discussion on disordered GNRs of fixed length and fixed applied (time-independent) voltage. In Figure~\ref{fig:hopdisorder}, we introduce disorder into the GNR without breaking chiral symmetry. This is done by drawing a random number from a uniform distribution of width $w$ around the average hopping matrix value $t_C$. We exclude second and third nearest neighbor hoppings, and we consider randomness only in the hoppings, for we wish to preserve particle-hole symmetry~\cite{CastroNeto2009,Gopar2016}.
We see that the disorder appears to increase the traversal time and also has the effect of decreasing the quality of the signal, so that multiple side-peaks are visible. These are caused by internal reflections induced by the disorder. In addition, the intrinsic resonant frequencies for the disordered GNRs (shown by the Fourier spectra) are red-shifted due to the hopping disorder. This finding is consistent with the idea of reduced operational frequencies for the GNR devices due to disorder.

\begin{figure}[t]
\centering
\includegraphics[width=\textwidth]{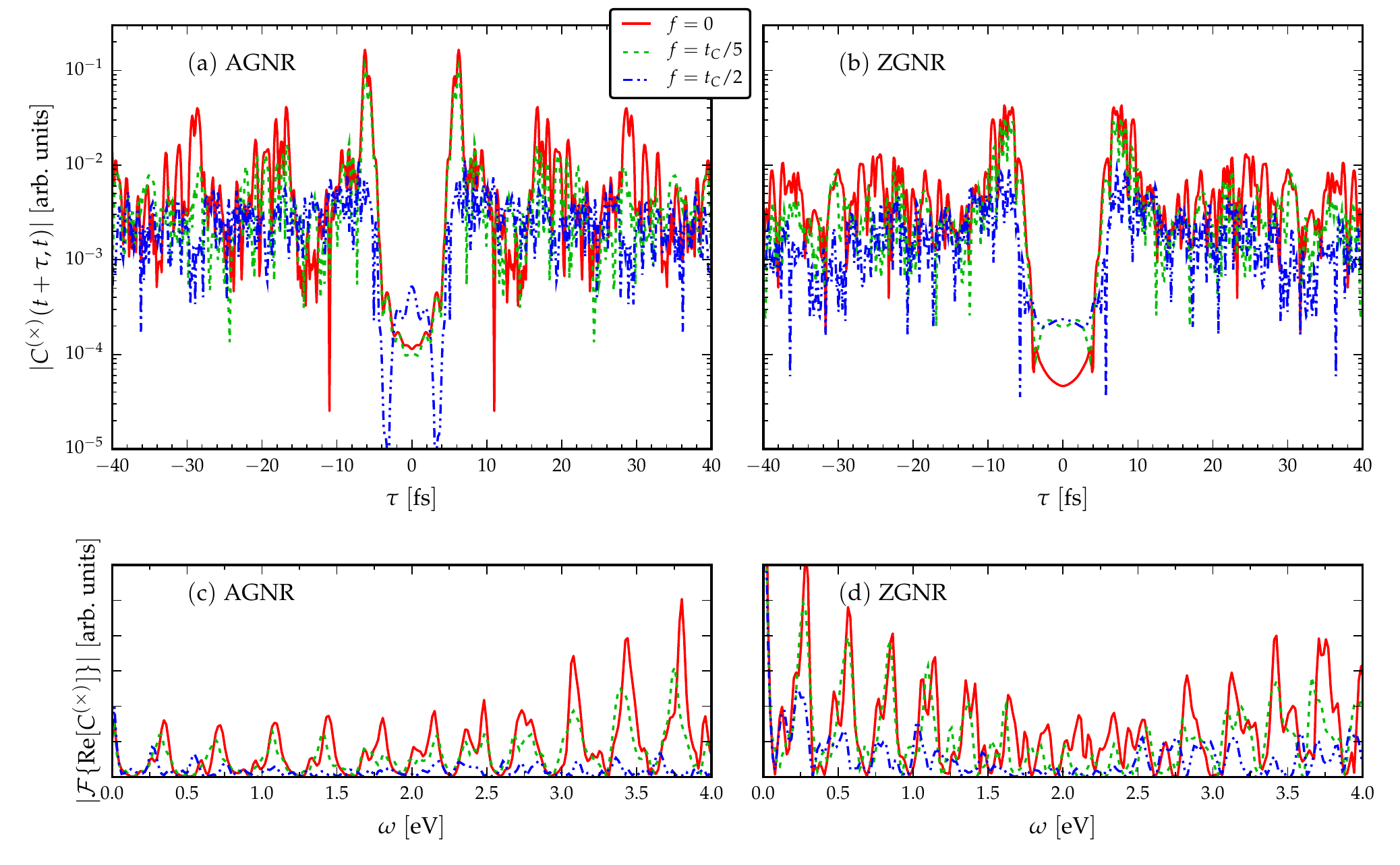}
\caption{Current cross-correlations for fixed length disordered GNRs ($L=5$~nm) with fixed voltage $V_L=-V_R=t_C/2$. A uniformly-distributed chiral symmetry-breaking disorder $f$ is included in the on-site energy.}
\label{fig:onsitedisorder}
\end{figure}

In Figure~\ref{fig:onsitedisorder}, we break the chiral symmetry by adding a random term to the on-site energy levels of the GNR. This is also done by drawing a random number from a uniform distribution of width $f$ around the zero on-site energy for the pristine GNR. In Figure~\ref{fig:onsitedisorder}(a), we look at the AGNR case, and we observe in this case the deterioration of a clear traversal time signal as the Anderson localisation increases the average dwell time in the interior of the GNR. In Figure~\ref{fig:onsitedisorder}(b), the average traversal time is once again seen to be larger in the ZGNR case. Interestingly, there is a crossover in both GNR configurations as the disorder destroys the coherence of the propagating wave packet around $f=t_C/2$. In contrast to the case of hopping disorder, here the operational frequencies of the GNR device (shown by the Fourier spectra) remain roughly unchanged for the on-site disorder. This finding could be related to the character of the disorder: While both types of disorder may introduce an effective tunnel barrier around the disordered GNR that the propagating electrons must overcome, the hopping disorder case corresponds to deformation of the lattice geometry whereas the on-site disorder corresponds to a change in charge neutrality or chemical potential.

\subsection{Response to an ac drive}

\begin{figure}[t]
\centering
\includegraphics[width=\textwidth]{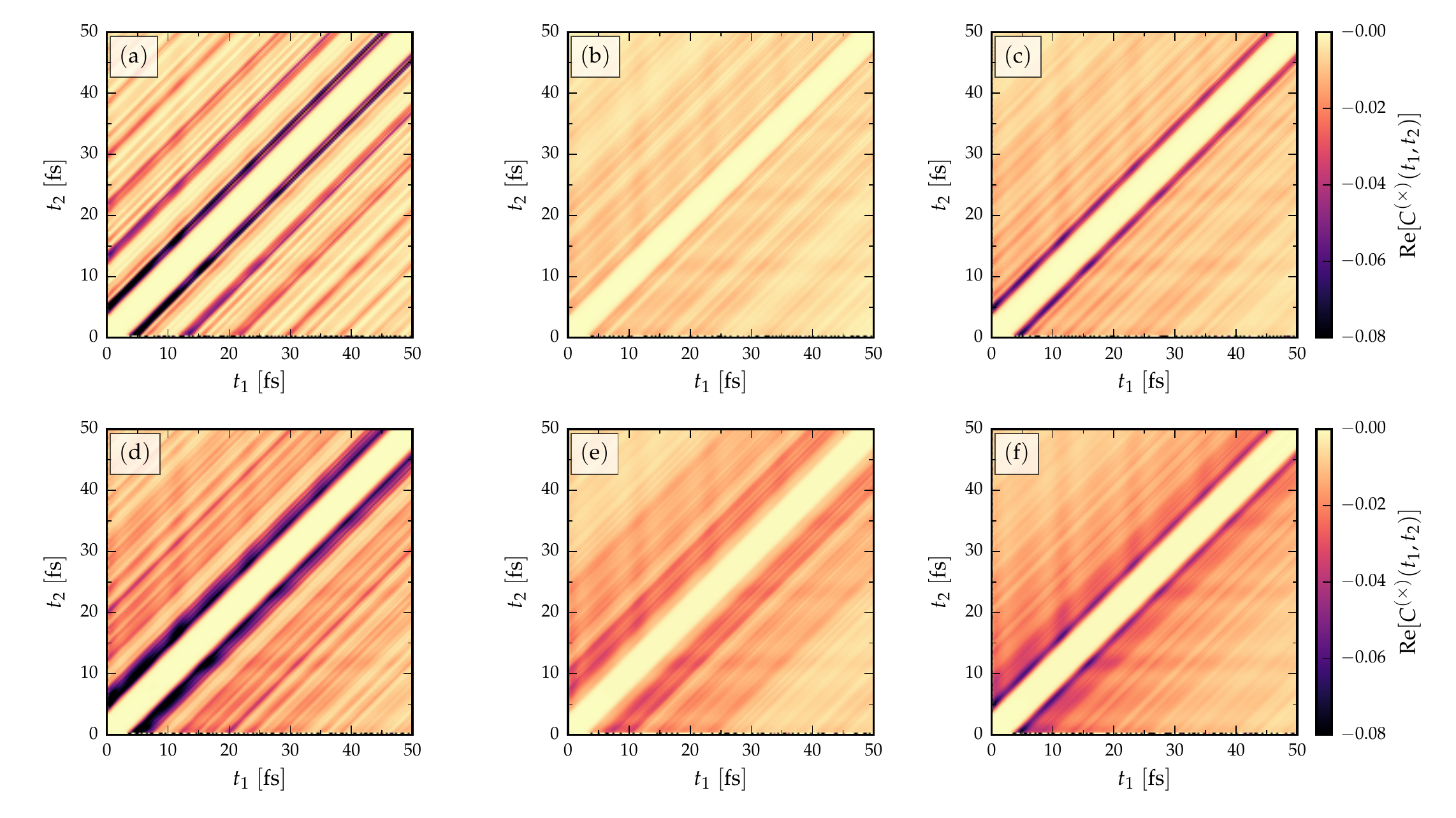}
\caption{Two-time current cross correlations for fixed length ($L=2$~nm) GNRs with ac driving. (a) AGNR without disorder, (b) AGNR with hopping disorder $w=t_C/2$, (c) AGNR with on-site disorder $f=t_C/2$, (d) ZGNR without disorder, (e) ZGNR with hopping disorder $w=t_C/2$, (f) ZGNR with on-site disorder $f=t_C/2$.}
\label{fig:2dplots}
\end{figure}

We finally address the full two-time character of the cross correlation~\eqref{eq:correlator}. Specifically, we consider the case of ac driving by introducing a monoharmonic driving term to the voltage
\begin{equation}
V(t) = V_0 + A \cos(\varOmega t)
\end{equation}
where the static part is set by $V_0=t_C/2$ and the amplitude of the ac driving is $A=t_C/2$ with the driving frequency $\varOmega = t_C/10$. In order to reduce the computational effort, we consider only the short nanoribbons in this case ($L=2$~nm). In Figure~\ref{fig:2dplots} we show the propagation of the full two-time cross correlation from the initial time $t_{0}$. In contrast to the previous steady state results, here we show the initial transient (up to $50$~fs), which includes relaxation effects. 

We observe that the ac driving does not change the overall picture of traversal time, i.e., the time it takes for the information to traverse through the nanoribbons can be clearly read off from the separation of peaks along the anti-diagonal. Compared to the long-time limit in Figures~\ref{fig:nodisorder},~\ref{fig:hopdisorder}, and~\ref{fig:onsitedisorder}, the initial transient only shows some additional oscillations but the main features seen in the steady state data are still visible. The two-time correlations also show the effect of disorder; as in the dc case, the signal gets considerably disturbed for hopping disorder (cf. Figure~\ref{fig:hopdisorder}) and for on-site disorder (cf. Figure~\ref{fig:onsitedisorder}), but in the latter case the signal destruction is less severe. In calculations not shown here, we have checked that other types of ac driving (biharmonic drive, faster/slower modulation, lower/higher intensity) have no effect on the qualitative behaviour of traversal times.

\section{Conclusions}

\begin{figure}[t]
\centering
\includegraphics[width=\textwidth]{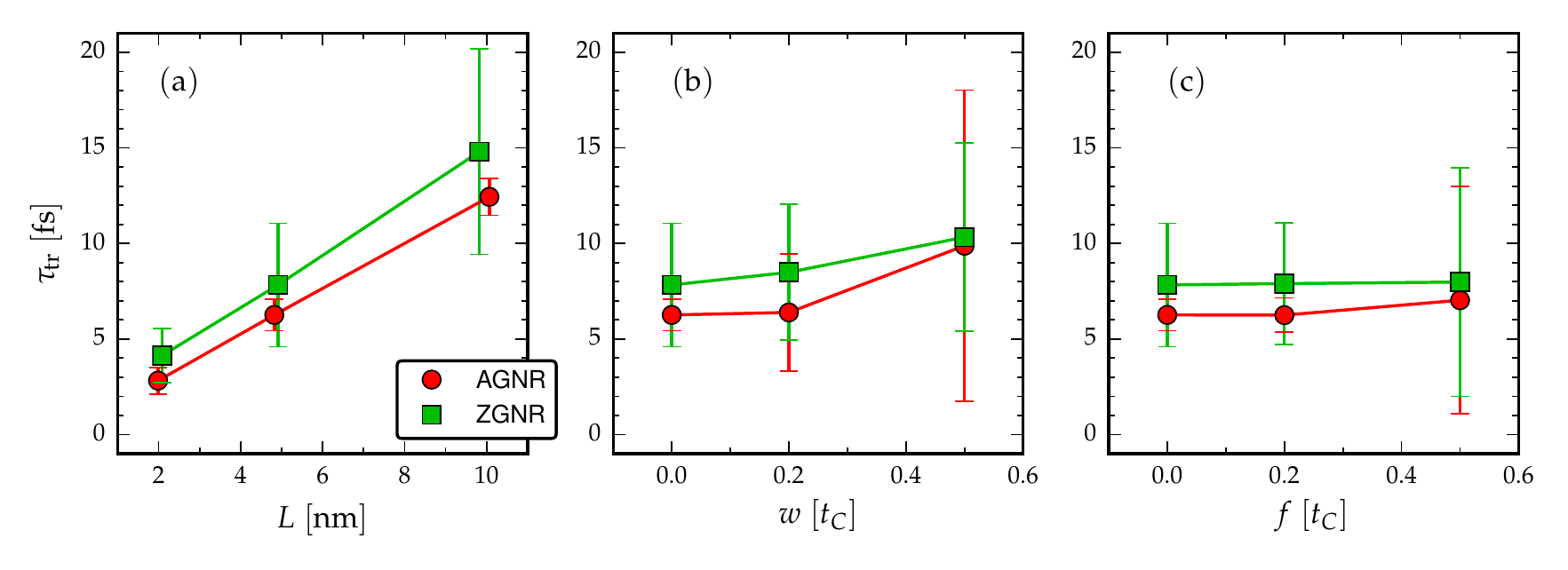}
\caption{Electron traversal times estimated from the distance between the first maxima in the current cross correlation (cf. Eq.~\eqref{eq:ttime_def} and Fig.~\ref{fig:nodisorder}). (a) No disorder, fixed voltage $V_L=-V_R=t_C/2$, varying length $L$. (b) Fixed length $L=5$~nm, fixed voltage $V_L=-V_R=t_C/2$, varying hopping disorder $w$. (c) Fixed length $L=5$~nm, fixed voltage $V_L=-V_R=t_C/2$, varying on-site disorder $f$. The error bars are empirically estimated from the cross correlation peaks as the full width at half maximum.}
\label{fig:tautr}
\end{figure}

We have employed the recently developed TD-LB formalism to compute the two-time current correlation functions for disordered GNRs. This methodology is a fast and accurate way of addressing mesoscopic quantum transport phenomena out of equilibrium as it is well-supported by the underlying nonequilibrium Green's function theory~\cite{svlbook}. By our analysis we confirm that the current cross-correlation is a good measure of electron traversal time. We find that the traversal time scales roughly linearly with the length of the GNRs, and that the traversal time also depends strongly on the GNR orientation.

We found that disorder in GNRs increases the traversal times, in general, and ultimately destroys the whole picture of coherent information transfer over the GNR junction when the disorder-induced scattering is strong. The ``rule-of-thumb'' character of our findings is summarized in Figure~\ref{fig:tautr}. We considered two types of disorder, one that preserves (hopping disorder) and one that breaks (on-site disorder) the chiral symmetry of the GNR. In Fig.~\ref{fig:tautr}(b) we find that the intrinsic operational frequency of the GNR is redshifted for the hopping disorder while in Fig.~\ref{fig:tautr}(c) we see that it remains roughly unchanged for the on-site disorder. However, in the latter case the statistical spread of $\tau_{\text{tr}}$ is significantly enhanced as the on-site disorder is increased. To measure the current cross-correlation and extract experimental values for $\tau_{\text{tr}}$, there exist a range of spectroscopic techniques which relate the field strength of photons emitted from each lead to the current. The zero- and finite-frequency current cross-correlations can then be extracted from these current measurements~\cite{deng2015coupling,fevrier2018}.

Our noninteracting approach is sufficient for graphene structures since monolayer graphene devices have been experimentally shown to have ballistic transfer lengths on the order of hundreds of nanometers at low temperatures~\cite{Miao2007,Lin2009}. Even though our approach is limited to noninteracting electrons, we expect the current correlations and traversal times to be similarly related even when dealing with, e.g., electron-electron or electron-phonon interactions~\cite{Galperin2007,Swenson2012,Haertle2013,Ridley2019}. If perturbation theory could be applied, i.e., when the interaction is weak, current correlation or noise simulations are still feasible to perform in terms of the one-particle Green's function~\cite{Galperin2006,Souza2008,Myohanen2009,Lynn2016,Miwa2017,Cabra2018}. Disordered interacting systems have also been studied using mean-field or density-functional theories~\cite{Shepelyansky1994,Vojta1998,Karlsson2018}. Here out-of-equilibrium dynamics only due to voltage bias was considered but also thermal gradients could be included similarly~\cite{Eich2014a,Eich2014b,Eich2016a,Eich2016b,Covito2018}. At present, for the case of strong interaction these approaches cannot yet be extended to realistic device structures since considerably more complicated and numerically expensive methods are required~\cite{Ridley2018b}.

We also confirmed that the overall picture of electron traversal times is not qualitatively changed by introducing an ac driving voltage compared to the response to a dc drive. Possible quantitative differences in the response signals to an ac drive could be related to signatures of photon assisted tunneling on traversal time~\cite{Ridley2017b,Tuovinen2019b} but, for now, will be left for future work. On the other hand, it would also be interesting to consider, e.g., a short laser pulse for exciting the system out of equilibrium~\cite{Kemper2015,Sentef2016,Kemper2017} instead of the quench of the voltage bias employed in the present work. These topics will also be addressed more thoroughly in a forthcoming paper.


\acknowledgments

We wish to thank Robert van Leeuwen for productive discussions. This research was funded by the Raymond and Beverly Sackler Center for Computational Molecular and Materials Science, Tel Aviv University (M.R.) and by the DFG Grant No. SE 2558/2-1 through the Emmy Noether program (M.A.S. and R.T.).


%


\end{document}